# High thermal conductivity in semiconducting Janus and non-Janus diamanes

Mostafa Raeisi[#a], Bohayra Mortazavi*[,#,b], Evgeny V. Podryabinkin[c], Fazel Shojaei[d], Xiaoying Zhuang**[,b] and Alexander V. Shapeev[c]

[a]Department of Mechanical Engineering, Imam Khomeini International University, Qazvin, Iran
[b]Chair of Computational Science and Simulation Technology, Department of Mathematics and Physics, Leibniz University Hannover, Germany, Appelstraße 11, 30157 Hannover, Germany.
[c]Skolkovo Institute of Science and Technology, Skolkovo Innovation Center, Nobel St. 3, Moscow 143026, Russia.
[d]Department of Chemistry, College of Sciences, Persian Gulf University, Boushehr 75168, Iran.

**Abstract**

Most recently, F-diamane monolayer was experimentally realized by the fluorination of bilayer graphene. In this work we elaborately explore the electronic and thermal conductivity responses of diamane lattices with homo or hetero functional groups, including: non-Janus $C_2H$, $C_2F$ and $C_2Cl$ diamane and Janus counterparts of $C_4HF$, $C_4HCl$ and $C_4FCl$. Noticeably, $C_2H$, $C_2F$, $C_2Cl$, $C_4HF$, $C_4HCl$ and $C_4FCl$ diamanes are found to show electronic diverse band gaps of, 3.86, 5.68, 2.42, 4.17, 0.86, and 2.05 eV, on the basis of HSE06 method estimations. The thermal conductivity of diamane nanosheets was acquired using the full iterative solutions of the Boltzmann transport equation, with substantially accelerated calculations by employing machine-learning interatomic potentials in obtaining the anharmonic force constants. According to our results, the room temperature lattice thermal conductivity of graphene and $C_2H$, $C_2F$, $C_2Cl$, $C_4HF$, $C_4HCl$ and $C_4FCl$ diamane monolayers are estimated to be 3636, 1145, 377, 146, 454, 244 and 196 W/mK, respectively. The underlying mechanisms resulting in significant effects of functional groups on the thermal conductivity of diamane nanosheets were thoroughly explored. Our results highlight the substantial role of functional groups on the electronic and thermal conduction responses of diamane nanosheets.

## 1. Introduction

Graphene [1–3], owing to its exceptional properties, succeeded in the evolution of two-dimensional (2D) materials as one of the most important classes of materials. 2D materials are currently considered as one of the most appealing nanomaterials to improve the design and push the boundaries in various critical technologies, such as, nanoelectronics, nanophotonics, and energy conversion/storage systems. The outstanding properties of the graphene, motivated the design and synthesis of a wide variety of other 2D counterparts, among them $MoS_2$ [4,5] and phosphorene [6,7] have attracted remarkable attention due to their inherent narrow electronic band gaps. The zero band gap electronic response of graphene is a drawback for various applications, particularly in nanoelectronics. But, on the other hand, it has nonetheless played a positive role in stimulating the design and fabrication of 2D inherent semiconductors.

In comparison with graphene, other carbon-free 2D materials are known to exhibit lower thermal conductivities and mechanical properties. Therefore, designing novel 2D materials that show electronic band gaps and simultaneously exhibit comparable thermal conductivity and tensile stiffness to those of graphene is yet among the most appealing fields of research in 2D materials. One of the promising classes of such materials is carbon-based compositions. For example, polyaniline $C_3N$ introduced experimentally by Mahmood *et al.* [8] in 2016 is a graphene-like and carbon-based 2D material, which is a semiconductor with remarkably high stiffness and thermal conductivity [9–14]. Nitrogenated holey graphene, $C_2N$, is another carbon-based 2D semiconductor which was synthesized in 2015 [15], but its nanoporous lattice results in two orders of magnitude lower thermal conductivity than graphene [14,16–18]. It is noticeable that in the case of polyaniline $C_3N$, has graphene-like structure and ordered arrangement of nitrogen atoms; however, its thermal conductivity is by order of magnitude lower than that of the graphene.

From ab-initio simulations [19–21], it has been already known that by converting the bilayer graphene into a diamond monolayer, the so-called "diamane", it is possible to open a band gap. Due to the presence of strong covalent bonding between carbon atoms, the diamane nanosheets are expected to be mechanically strong and show high thermal conductivity. In a recent theoretical study [22], it was notably predicted that H-diamane can show a close thermal conductivity to that of the hydrogenated graphene. From the theoretical point of view, the hybridization between $sp^3$ orbitals and metal surface $d_z^2$ orbitals are the main driving force for the conversion of bilayer graphene to diamane [23]. Therefore, the outer surfaces of original graphene nanomembranes are required to be functionalized with fluorine, hydrogen, or hydroxyl groups to accomplish this conversion and also for the stability of fabricated diamane nanosheets. In fact by removal of functional atoms, the full-carbon diamane does not stay stable and converts to bilayer graphene. The theoretical concept of diamane nanomembranes was most recently experimentally approved by Bakharev and coworkers [24]. They show that fluorine chemisorption on bilayer graphene grown by chemical vapour deposition (CVD) can facilitate the synthesis of F-diamane. This experimental accomplishment is expected to pave the path for the realization of other diamane counterparts, synthesizable from hydrogenated or chlorinated graphene. In addition, there might be a possibility that Janus diamane nanomembranes could be also fabricated, in which the outer surfaces of bilayer graphene are functionalized with hetero atoms. Worthy to note that in recent studies, fabrication mechanism of diamane nanosheets [25–27], and the application of H-diamane as a mechanical resonator [28] have been explored. However, a complete overview on the dynamical stability, electronic nature and thermal conductivity of Janus and non-Janus diamanes has not been established.

In this work, we first perform first-principles simulations to examine the dynamical and elastic stability of $C_2H$, $C_2F$, $C_2Cl$, $C_4HF$, $C_4HCl$ and $C_4FCl$ diamane nanosheets, as novel carbon based 2D systems. Next, we investigate the electronic nature of aforementioned nanosheets by employing the HSE06 hybrid functional. Finally, we conduct a throughout exploration of thermal conductivity of these novel 2D systems, by iterative solutions of the Boltzmann transport equation. These calculations were substantially accelerated by employing machine-learning interatomic potentials in obtaining the third-order force constants, which are the most computationally demanding part for evaluating the thermal conductivity. The validity of employed approach was confirmed by observing a remarkable accuracy in the estimation of graphene's thermal conductivity, as compared with experimental measurements and pure DFT results. Our first-principles-based investigation not only provides a comprehensive vision on the stability and electronic responses of a novel class of carbon based 2D materials, but also highlights the substantial role of functional groups on the electronic character and thermal transport along diamane nanosheets. These extensive results on a novel class of 2D materials along with the employed highly computationally efficient numerical method are expected to serve as valuable findings for the further studies.

## 2. Computational methods

First-principles density functional theory (DFT) simulations in this work were carried out employing the *Vienna Ab-initio Simulation Package* (VASP) [29–31]. Generalized gradient approximation (GGA) with Perdew-Burke-Ernzerhof revised for solids (PBEsol) [32] were adopted. Plane-wave cutoff energy of 500 eV with a convergence criterion of $10^{-5}$ eV was used for the self-consistent electronic loop. The geometry optimization was conducted using the conjugate gradient method with the convergence criterion of 0.001 eV/Å for Hellmann-Feynman forces with a 20×20×1 Monkhorst-Pack [33] k-point mesh. Periodic boundary conditions were applied along with all three Cartesian directions with a fixed simulation box size of 20 Å along the normal-to-sheet direction, to avoid imaginary interactions. Because of the underestimation of the band gap by the PBE/GGA functional, the screened hybrid functional of HSE06 [34] was employed to more precisely evaluate the electronic band structure. Density functional perturbation theory (DFPT) simulations were carried out over 5×5×1 and 10×10×1 supercells, for graphene and diamane monolayers, respectively using 3×3×1 k-point mesh. PHONOPY package [35] was then employed to acquire the phonon dispersions and harmonic (second-order) force constants with the DFPT results as inputs.

In this work, we used moment tensor potentials (MTPs)[36] to replace the DFT calculations in the force constant calculations. We remind that MTP belongs to the machine-learning interatomic potentials [37–39] family, which accordingly offers a flexible functional form with high level of accuracy. MTPs have been successfully employed to predict novel materials [40,41], study lattice dynamics [42–44] and thermal conductivity [45,46] in various systems. In this work, we conducted the fitting of MTPs over short ab-initio molecular dynamics (AIMD) trajectories. In our latest study [42], we showed that MTPs trained over short AIMD trajectories can very accurately reproduce the phononic properties of complex 2D lattices as compared with DFPT simulations. In aforementioned study [42], the theoretical background and computational details for the practical employment MTPs in the evaluation of phononic properties are more elaborately discussed. In accordance with our recent study [42], in this work the training sets were also prepared by conducting the AIMD simulations within the PBEsol/GGA functional for 1000 time steps at 50 K, over 4×3×1 supercells using a 2×2×1 k-point grid with a time step of 1 fs. The lattice thermal conductivity and complex phononic properties were then acquired by the full iterative solution of the Boltzmann transport equation using the ShengBTE [47] package. In this approach, the most challenging and computationally demanding section is to acquire the anharmonic force constants, which normally requires several hundred DFT calculations over supercells. In this work, passively fitted MTPs replace the DFT calculations for the evaluation of anharmonic force constants. In this case we used the same supercells as those we used in our DFPT calculations, and the interactions with the ninths nearest neighbours were considered.

## 3. Results and discussion

With the PBEsol functional, the lattice constant of bulk diamond was found to be 3.556 Å, which is remarkably close to the experimentally reported value of 3.567 Å. The lattice constant of graphene was also found to be 2.461 Å, close to the experimentally measured value of ~2.465 Å [48]. In addition, using this functional the average interlayer separation of carbon layers in $C_2F$ diamane was found to be 2.048 Å, in outstanding agreement with the experimentally measured value of 2.05 Å [24]. With PBEsol the lattice constant of graphene was also found to be 2.461 Å, which is also very close to the reported value of 2.464 Å [48]. It is thus clear that PBEsol is an accurate method for describing the atomic structure in carbon-based materials. Top and side views of geometry optimized diamane monolayers considered in this work are shown in Fig. 1, demonstrating four-layer sandwiched structures with hexagonal atomic lattices. The hexagonal lattice constants of $C_2H$, $C_2F$, $C_2Cl$, $C_4HF$, $C_4HCl$ and $C_4FCl$ were found to be 2.517, 2.546, 2.734, 2.532, 2.640 and 2.651 Å, respectively. The energy minimized lattices are included in the supplementary information document. In Fig. 1 we also plotted the electron localization function (ELF) [49] for the side views to investigate the types of chemical bonds in these systems. As it is clear, a high degree of electron localization in the intermediate regions between each two directly bonded atoms with high iso-surface ELF values over 0.7, confirm the strong covalent bonding between C-C, C-H, C-F, and C-Cl bonds throughout the diamane nanosheets. Conspicuous electron localization occurs around the bonded halogen atoms, due to their lone pair electrons. We conducted the AIMD simulations at 500 K for 20 ps, and all the diamane monolayers were found to be completely intact, thus confirming the desirable thermal stability of these novel 2D systems.

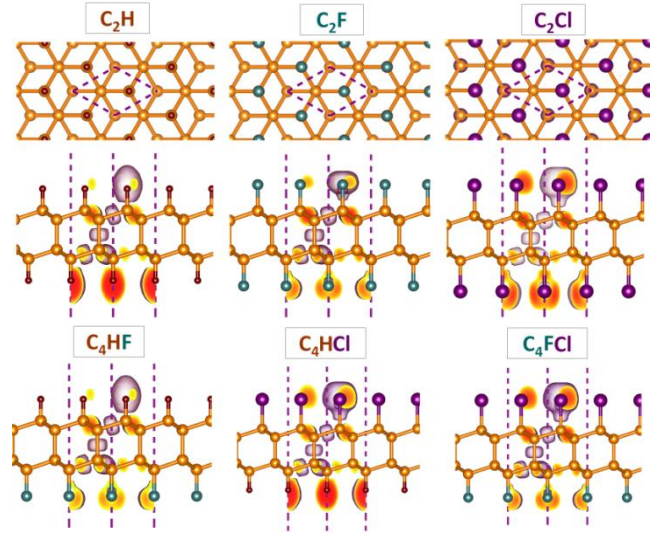

**Fig. 1**, Atomic structure of diamane nanosheets considered in this work. Iso-surfaces (set at 0.7) illustrate the electron localization function [49] within the unitcell.

In the next step, we examine the elastic modulus and electronic properties of these novel 2D systems. In this work, the thickness of diamane monolayers was defined by also including the van-der-Waals radii of the two functional groups to the apparent thickness. This way, the thickness of $C_2H$, $C_2F$, $C_2Cl$, $C_4HF$, $C_4HCl$ and $C_4FCl$ were found to be 6.95, 8.23, 9.5, 7.59, 8.22 and 8.55 Å, respectively. By applying uniaxial straining along the two different directions, we observed that these nanomembranes show isotropic elasticity. The elastic modulus of $C_2H$, $C_2F$, $C_2Cl$, $C_4HF$, $C_4HCl$ and $C_4FCl$ were estimated to be 692, 574, 430, 633, 550 and 522 GPa. These results interestingly show that the elastic modulus decreases by increasing the weight of functional groups, this way $C_2H$ and $C_2Cl$ diamane are respectively the strongest and weakest lattices. It was found that by increasing of the functional groups weight, the length of C-C bond connected with the functional atom increases, resulting in the slight softening of this bond. Because of their orientation toward the loading, these aforementioned bonds dominate the in-plane mechanical response. Moreover, as discussed earlier the thickness of the structure increases by replacing the H atoms with F and Cl atoms, and such that softer bonds and higher thickness simultaneously result in lower elastic modulus. In comparison with graphene, the elastic moduli of diamane monolayers are by between 30-50% lower.

We next investigate the electronic properties of diamane nanosheets by analyzing the electronic band structure using PBEsol and hybrid HSE06 functionals. For comparison, we also calculated the band gap of bulk diamond using the HSE06 functional, estimated to be 5.33 eV, very close to the experimentally measured value of 5.5 eV. As evident in Fig. 2, the diamane monolayers exhibit semiconducting features with direct HSE06 (PBEsol) band gaps of 3.86 (3.06), 5.68 (3.95), 2.42 (1.24), 4.17 (3.12), 0.86 (0.16), and 2.05 (0.85) eV for $C_2H$, $C_2F$, $C_2Cl$, $C_4HF$, $C_4HCl$ and $C_4FCl$ monolayers, respectively. For these monolayers, the valance band maximum (VBM) and conduction band minimum (CBM) occur at Γ-point. Interestingly, despite some similarities between the band structures of Janus and non-Janus monolayers (e.g., band dispersions around the Fermi level), Janus structures exhibit different band gap values and absolute band edge positions. For example, the HSE06 calculated band gap for $C_4HCl$ monolayer (0.86 V) is found to be appreciably smaller than those of non-Janus parent $C_2H$ (3.86 eV) and $C_2Cl$ (2.42 eV) monolayers. We also observed a similar relationship between the band gap values of Janus $C_4FCl$ (2.05 eV) and non-Janus $C_2F$ (5.68 eV), $C_2Cl$ (2.42 eV) parents. Relatively small direct band gaps and highly dispersed valance and conductions bands, associated with very small charge carrier effective masses, makes Janus $C_4HCl$ and $C_4FCl$ monolayers promising materials for applications in nanoelectronics and optoelectronic devices.

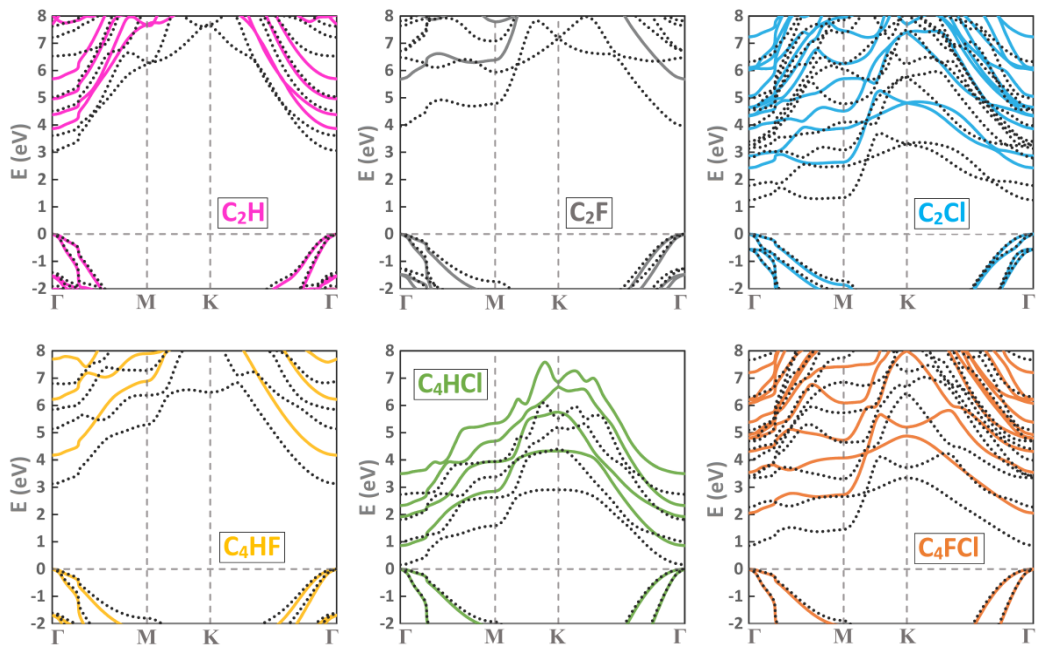

**Fig. 2**, Electronic band structures of diamane monolayer, computed using PBEsol (dotted lines) and hybrid HSE06 (continuous lines) methods. Valance band maximum is set to zero energy.

Phonon dispersion of diamane structures are illustrated in Fig. 3, where DFT results are compared with those predicted by the MTPs method. It should be mentioned that hydrogenated diamane structures ($C_2H$, $C_4HCl$, and $C_4HF$) have two distant modes (approximately at 86 THz), which are not displayed in Fig. 3. The obtained results show a considerably close agreement between MTPs and ab-initio results. In particular, MTPs yield more smooth curves near the Γ-point, for which in a few cases the DFT results show marginal imaginary frequencies [42]. The quadratic dispersion characteristic of 2D materials is also obvious for flexural out-of-plane acoustical (ZA) branches of these structures. By comparing the Γ-M and Γ-K path near to the Γ-point, it is clear that all structures show similar behavior, which indicates isotropic properties. For the simplification of our analysis of thermal transport, optical modes are divided into two portions: lower optical modes (LOM) and upper optical modes (UOM). For example, $C_2H$ has three LOM, whereas $C_2F$, $C_2Cl$, and $C_4FCl$ have seven lower optical modes, and finally, $C_4HF$ and $C_4HCl$ have five lower optical modes. Here, two mechanisms control phonon transport and, consequently, thermal conductivity. First, showing wide or narrow dispersion of each mode, specifically acoustical ones, indicate higher or smaller group velocity, respectfully. As it is evident, acoustic branches of $C_2H$ are wider in comparison to other diamane counterparts. $C_4HF$ and $C_4HCl$ are two next structures with wider frequency ranges due to light H atoms and unevenly atomic masses [50], which also leads to higher group velocities. For the studied structures, the dispersions of LOMs are also clearly wider than UOMs. Second, exhibiting lower maximum frequency, like that of the $C_2Cl$ monolayer, usually results in the increase of the scattering and consequently declined thermal conductivity. Generally, branches that show narrow and close dispersions, increase the scattering probability, and subsequently results in lower thermal conductivity. In addition to the narrower acoustical branches in $C_2F$, $C_2Cl$, and $C_4FCl$ structures, acoustic and LOM branches also cross each other, increasing the phonon scattering and thereby suppressing the thermal conductivity. $C_2H$ shows softened acoustic branches without interfering with other modes. In this line, total phase space (fraction of allowed scattering) of $C_2Cl$, $C_2F$, and $C_2H$ are, predicted to be $2.08 \times 10^{-3}$, $1.41 \times 10^{-3}$, and $5.79 \times 10^{-4}$, respectively and for non-Janus structures of $C_4FCl$, $C_4HCl$, and $C_4HF$ are $1.78 \times 10^{-3}$, $1.26 \times 10^{-3}$, and $9.94 \times 10^{-4}$, respectively, in which $C_2H$ shows the smallest values. Frequency of room temperature was obtained from: $\theta_\nu = \frac{h\nu}{K_\nu} = 300\ K \rightarrow \nu = 6.25\ THz$. According to the Bose-Einstein distribution function at room temperature (find Fig. S1), the probability of stimulated phonons with larger frequency than 6.25 THz is only ~8% and those larger than 10 THz is ~4%. Therefore, $C_2Cl$ structure exhibits more phonon modes interruptions and scattering among the other diamane structures.

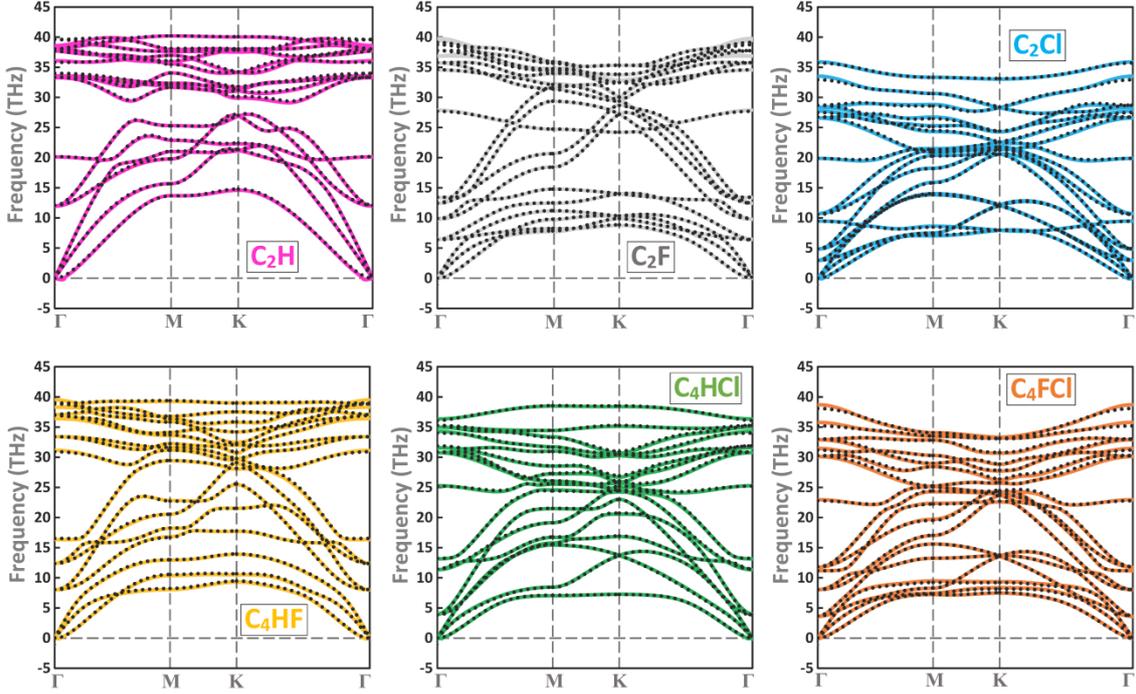

Fig. 3, Phonon dispersion of diamane monolayer, computed using DFT (dotted lines) and MTP (continuous lines) method. For the samples functionalized with H atom there exist almost flat bands at frequencies around 86 THz, which are not shown here.

MTP-based estimates for the thermal conductivities of graphene and diamanes at different temperature are compared in Fig. 4. In this case, temperature power factor (α), group velocity for the LA mode (near the Γ-point), estimated thermal conductivities at room temperature and Debye temperature are also compared in Fig. 4. The thermal conductivity of graphene is predicted to be as 3636 W/mK, which falls within the experimentally measured values of 1500-5300 W/mK [51–54]. In Fig. S2 we compared our MTP-based results for the temperature-dependent thermal conductivity of graphene with several experimental [55,56] and DFT based studies [57,58], which reveal close agreement. Fugallo et al. [57] calculated the thermal conductivity of graphene as high as 3600 W/mK, considerably close to our results. Furthermore, our MTP-based approach predicts a temperature power dependence in thermal conductivity, α ($K \sim T^{-\alpha}$) of 1.35 for graphene, close to DFT results of 1.32 by Lindsay *et al.* [58] and 1.34 by Fugallo *et al.* [57]. As it is well known, acoustic phonons are the main heat carriers in the graphene. In Fig. S3 we compare the contribution of three acoustic modes on the overall thermal conductivity by MTP and previous theoretical studies. These comparisons clearly highlight the remarkable accuracy of our employed technique for estimating the lattice thermal conductivity. According to our results, the room temperature thermal conductivity of $C_2H$, $C_2F$, $C_2Cl$, $C_4HF$, $C_4HCl$ and $C_4FCl$ diamane monolayers are estimated to be 1145, 377, 146, 454, 244 and 196 W/mK, respectively. Previously, Zhu *et al.* [22] calculated the thermal conductivity of $C_2H$ as high as ~1960 W/mK, which was found to be comparable to that of the hydrogenated graphene. Zhu and coworkers [59] recently reported the thermal conductivity of $C_2F$ to be 362 W/mK, which is remarkably close to our result. According to our results for the temperature power factor, α, $C_2H$ and $C_2Cl$ show, respectively the highest and lowest power factor, among the considered structures. At low temperatures, a small portion of low-energy phonons stimulate, propagate, and scatter through the crystal, which decreases the Umklapp (U) scattering process, for which the momentum is not conserved. Hence, the normal (N) scattering process is dominant at low temperatures. Having a larger α in $C_2H$ indicates its larger U scattering and consequently thermal resistance by increasing temperature. $C_4HCl$ Janus diamane structure shows a higher thermal conductivity than non-Janus $C_2Cl$. Similarly, Janus $C_4HF$ also yields a lower and higher thermal conductivity than its non-Janus $C_2H$ and $C_2F$ counterparts, respectively. It is interesting that unlike the electronic properties the thermal conductivity of Janus diamanes fall in between of their non-Janus counterparts.

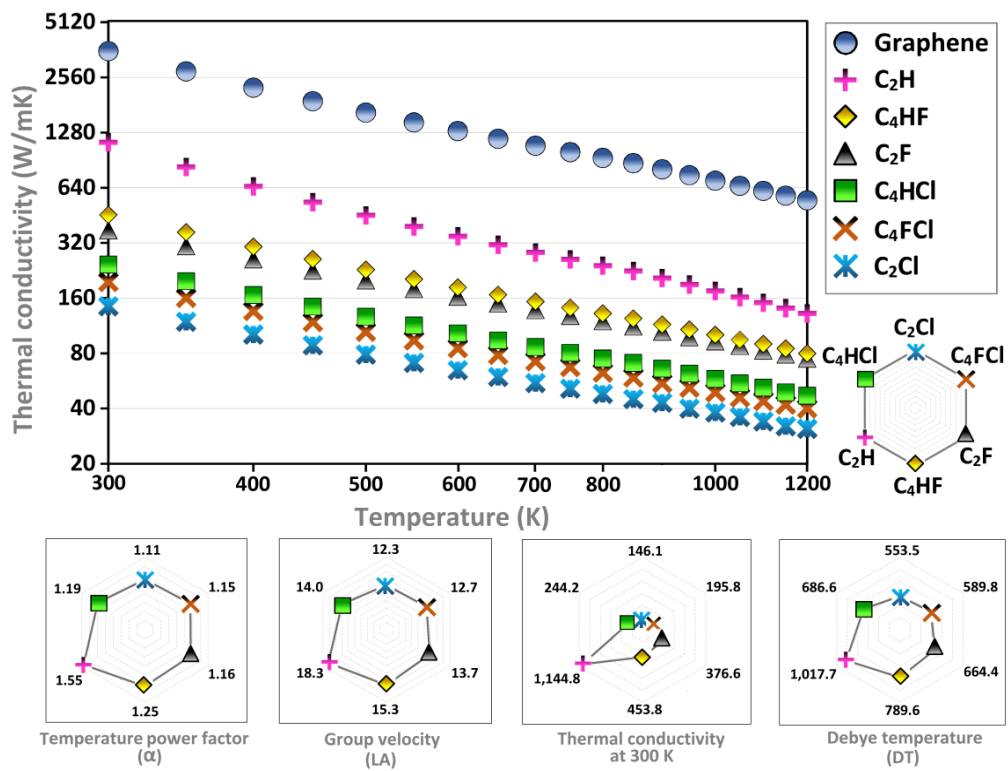

**Fig. 4**, Estimated lattice thermal conductivities as a function of temperature. The lower panels illustrate the temperature power factor, the group velocity of LA modes (km/s), thermal conductivity at room temperature (W/mK) and Debye temperature (K), respectively.

Thermal conductivity contribution for each phonon mode is depicted in Fig. 5. Flexural acoustical (ZA) phonon shows 85.2% contribution to the graphene's thermal conductivity, which is in close agreement with previous reports (Fig. S3). Among all six diamane structures, $C_2H$ shows the highest contribution of ZA mode (36.6%). We remind that diamane presents a four atomic layered structure and such that is not like the graphene with fully planar atomic lattice, therefore unlike the graphene (having the selection rule) the ZA mode does not dominate the thermal transport in these systems. Because of the lighter weight of H atoms and its modest electronegativity in comparison with F and Cl counterparts, $C_2H$ exhibits the closest lattice to that of the graphene. Therefore, for the $C_2H$ diamane the sum of acoustic phonon modes show a remarkable 92.3% contribution to the total thermal conductivity. In the earlier study by Zhu *et al.* [22], they reported the contribution of ZA, TA, and LA bands as 40%, 30%, and 20%, respectively while, in the present work, we found corresponding values of 36.6%, 28.8%, and 26.9%, respectively. For the rest of diamane nanosheets, due to heavier functional groups for non-Janus $C_2F$ and $C_2Cl$ or mass disparity of Janus $C_4HF$, $C_4HCl$ and $C_4FCl$ structures, optical modes contribution on the total thermal conductivity enhances. According to our results LOMs contribute higher than UOMs counterparts to the thermal conductivity, due to Bose-Einstein distribution. As a general trend, an increase in the atomic weight of functional groups in diamane structures noticeably raises the contribution of optical phonons.

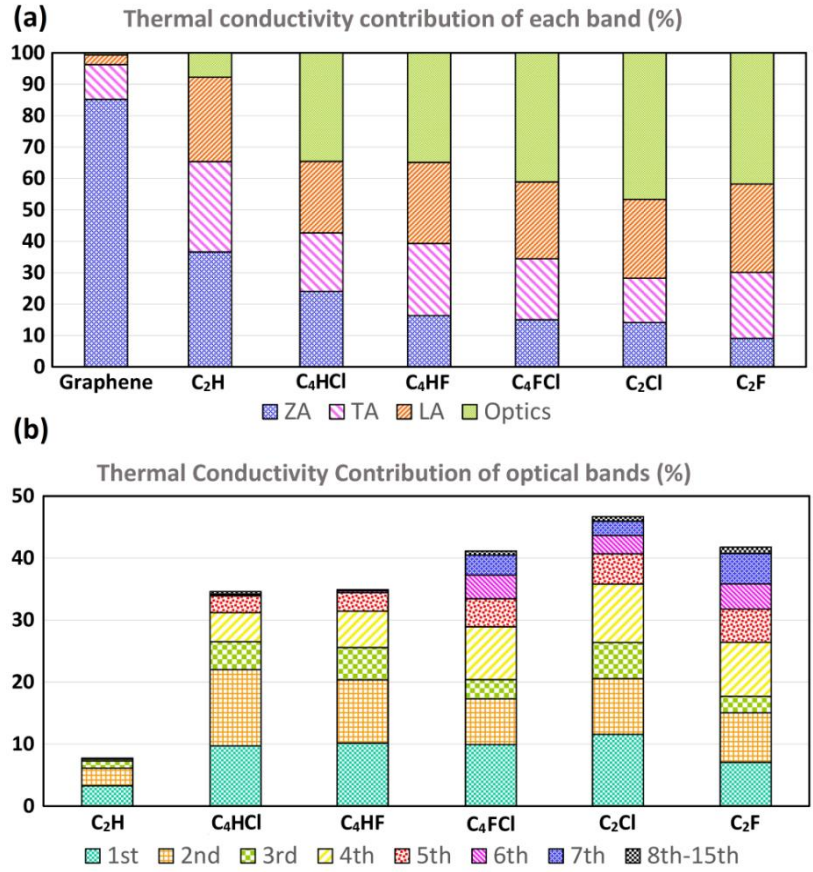

Fig. 5, Contributions of phonons mode on the total thermal conductivity. (a) Contribution of three main acoustic modes and all optical modes on the total thermal conductivity. (b) Contribution of different optical modes on the total thermal conductivity.

The accumulated thermal conductivities as a function of the sample mean free path are plotted in Fig. 6. $C_2H$ diamane shows two orders of magnitude larger effective mean free path (MFP) in comparison to other diamane structures. For this structure, the effective MFP stretches equivalently from 100 nm to 10 µm while, for the rest of the structures, the effective MFP is lower than 100-200 nm. As it is shown in Fig. S7, $C_2H$ shows smooth progress in the quantity of scattering rate within two orders of magnitude, i.e. range of 1 GHz to 100 GHz, for acoustic modes (energy less than 20 THz), which contributes to ~93% of the thermal conductivity. Having an equal width range of effective MFP, as that of the $C_2H$, could be explained by the scattering rate. Next, we explore the cumulative thermal conductivity as a function of phonon's frequency, as illustrated in Fig. S5. The start point of optical modes and the total contribution of acoustic modes are pinpointed in this Fig. S5b. It is conspicuous that frequencies lower than 20 THz dominate the thermal conductivity of diamane lattices. Moreover, lower frequencies have a larger effect on the thermal conductivity of the chlorine structures of $C_2Cl$, $C_4ClH$, and $C_4ClF$ in comparison with the other diamane lattices. It is noticeable that C$_4$HF shows an almost linear relation for cumulative function for the frequencies lower than 15 THz, which represents the wide range of frequencies contributed evenly to the thermal conductivity. The results shown in Fig. S5 also explains our previous discussions for phonon dispersion relations. For example, for the chlorine diamane structures, the start points of LOM (vertical line in the figure) are close together with low-energy (less than 5 THz), among which C$_2$Cl has shown a return point. After this return point, cumulative thermal conductivity of C$_2$Cl increases in a steep manner, and takes higher values than two other chlorine structures. If cumulative thermal conductivity shows a steep function and approaches to 100% in a low frequency, this implies that the thermal conductivity is dominated by low-frequency phonons, which limits the increase in the thermal conductivity. This suppression of thermal conductivity is due to having narrower band width (with the large number of crossing bands) for C$_2$Cl. In contrast, for C$_2$H structure, the starting point of LOM occurs at a higher frequency in comparison with other diamane structures.

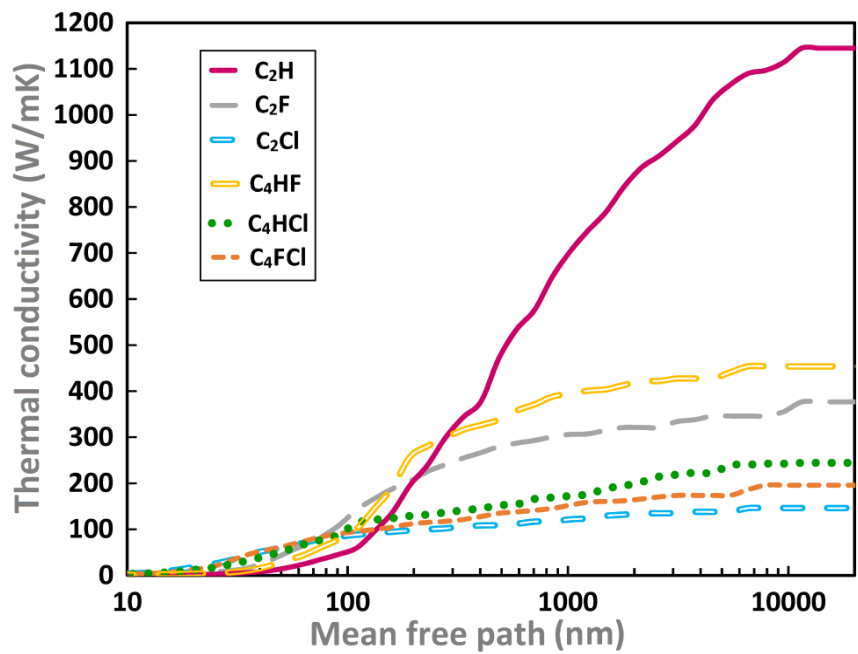

**Fig. 6**, Cumulative thermal conductivity of diamane nanosheets as a function of mean free path.

In order to study more deeply the thermal transport of diamane lattices, four major parameters including the phonon group velocity, scattering rates, Grüneisen parameter and weighted phase space were taken into consideration. The detailed information on these parameters as a function of frequency are illustrated, respectively, in Fig. 7 for the group velocity and Fig. S7 to Fig. S9 for the other three parameters. For graphene, the aforementioned properties acquired by the MTP-based method are in close agreement with previous DFT calculations, as illustrated in Fig. S6. For example, group velocity of graphene for LA and TA, is calculated to be 21.6 and 13.5 Km/s, respectively, near to the Γ-point, in agreement to previous DFT based solutions [60,61]. It is clear that the group velocity of $C_2H$ is highest among the diamane structures, which directly contribute to a higher thermal conductivity. Meanwhile, $C_2H$ has a more width-dedicated frequency (more than 10 THz) for acoustic bands in comparison with the other diamane counterparts, which further illustrates the larger contribution of acoustic phonons to the thermal conductivity. Comparing with the graphene's group velocity, for $C_2H$ diamane acoustic modes with high group velocities are suppressed at higher frequency, resulting in a lower thermal conductivity. For the rest of the diamane structures, the low frequency optical modes yield group velocity as high as that of LA band near the Γ-point, which can explain the higher contribution of optical modes to the thermal conductivity.

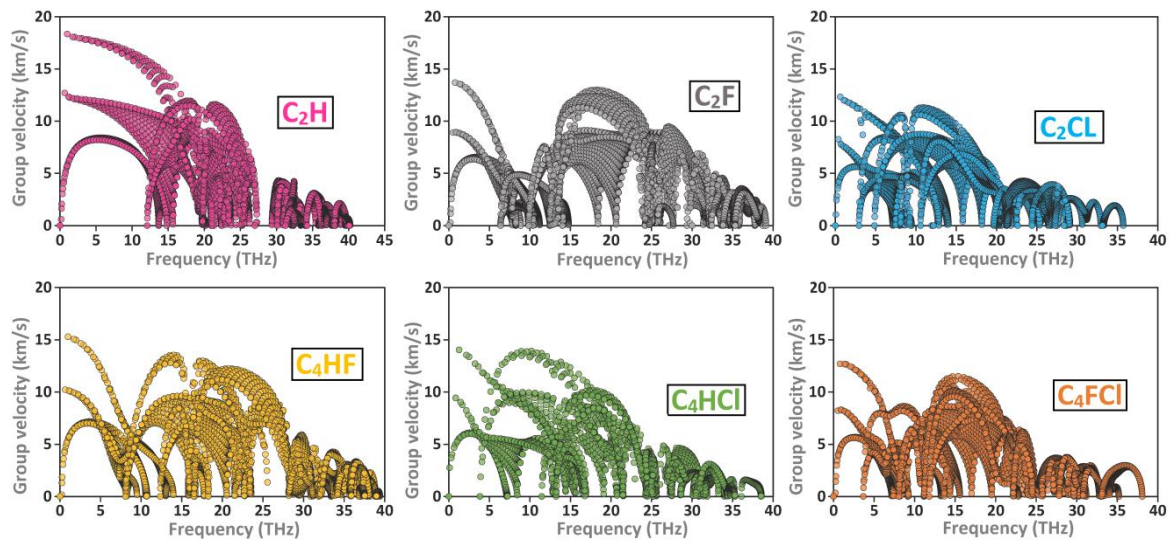

Fig. 7, Phonon group velocity as a function of frequency for diamane structures.

The scattering rates of diamane structures are compared in Fig. S7, which reveals that diamane lattices show similar behavior with values in the same range. With the increasing of frequency, the scattering rate increases monotonically. For $C_2H$, near the Γ-point and acoustic modes, the scattering rate is of lower magnitude with smother increase in comparison to the other diamane nanosheets, which can further explain its higher thermal conductivity and substantial contribution of acoustic modes. In this structure, LA, TA, and ZA coincide in all frequency range, which is not the case for graphene and other diamane structures. Comparing the graphene's scattering rate with those of diamane lattices, the former nonetheless shows higher scattering rate. In the case of graphene, the influence of its higher group velocity result in higher thermal conductivity as the thermal conductivity is proportional to the group velocity with power two (K=∑nv²τ).

The next influencing parameter is the Grüneisen parameter (γ), which is compared in Fig. S8. The absolute magnitude of the Grüneisen parameter has an inverse correlation with thermal conductivity. A larger value of this parameter implies higher anharmonicities of phonons, which could be interpreted as a stronger three-phonon scattering process. According to the previous works, [62,63], it is understandable that normal and Umklapp processes vary linearly and quadratically with frequency (i.e. ω and ω², respectively), respectively, according to Umklapp scattering equation:

$$\frac{1}{\tau_U} = 2\gamma^2 \frac{k_B T}{\mu V_0} \frac{\omega^2}{\omega_D} \qquad (1)$$

Here, $\tau_U$, $\gamma$, $\mu$, $\omega$, and $\omega_D$ are, respectively, the U process lifetime, Grüneisen parameter, shear modulus, angular frequency and Debye angular frequency and $k_B$, $T$, and $V_0$ are the Boltzmann constant, temperature, and volume per atom. For $C_2H$, ZA mode shows larger γ, which results in decreasing thermal conductivity. Meanwhile, the chlorine diamane structures ($C_2Cl$, $C_4HCl$, and $C_4FCl$) have high magnitude of γ in LOM, which stretched well and accompanied by ZA mode resulted in their lower thermal conductivity contribution. Comparing γ for $C_2H$ with the other diamane nanosheets shows a rather unexpected behavior, as its Grüneisen parameter suggest that it should yield a lower thermal conductivity.

The last parameter that should be taken into our investigation is the weighted phase space (Fig. S9), which represents possible scattering channels, for which energy is conserved in the entire the phase space. In the three phonon scattering process, illustrated in Fig. S15 (b), scattering channels occur if energy and momentum are conserved for the both absorption (+) and emission (-) process in Eq. 2 and 3, respectively.

$$\omega_1 \pm \omega_2 - \omega_3 = 0, \qquad (2)$$
$$\vec{q_1} \pm \vec{q_2} - \vec{q_3} = \vec{K}, \qquad (3)$$

Here, ω and $\vec{q}$ are angular frequency and wave vector respectfully, $\vec{K}$ is the reciprocal lattice vector which make the scattering process normal if becomes zero and Umklapp if becomes non-zero. Having a larger scattering phase space leads to lower thermal conductivity due to its more scattering channels. In the following expression of the weighted phase space (Eq. 4) [47,64], λ and $P$ represent the mode (optics or acoustics) and branch (in q-grid of wave vector), respectively. ω and $f$ are angular frequency and Bose-Einstein distribution function as a function of phonon's angular frequency. δ function is Dirac delta which approximated by Gaussian function in ShengBTE for implementation in programming.

$$W_{\lambda^1}^{absorption} = \Sigma_{\lambda^2 P^3}(f_{\lambda^2} - f_{\lambda^3})\left(\frac{\delta(\omega_{\lambda^1}+\omega_{\lambda^2}-\omega_{\lambda^3})}{\omega_{\lambda^1}\omega_{\lambda^2}\omega_{\lambda^3}}\right) \quad (4)$$

$$W_{\lambda^1}^{emission} = \Sigma_{\lambda^2 P^3}(f_{\lambda^2} + f_{\lambda^3} + 1)\left(\frac{\delta(\omega_{\lambda^1}-\omega_{\lambda^2}-\omega_{\lambda^3})}{\omega_{\lambda^1}\omega_{\lambda^2}\omega_{\lambda^3}}\right) \quad (5)$$

Our results show that $C_2Cl$ exhibit two-orders higher weighted phase space for acoustic and LOM than the three Janus and $C_2F$ counterparts, which explains its lower thermal conductivity. In contrast, $C_2H$ shows one-order lower scattering phase space in comparison with the rest of diamane structures, which also explain its higher thermal conductivity.

In addition to these four parameters as a function of frequency, their mappings over the reciprocal lattice are plotted in Fig. 8. More phonon modes (up to 6$^{th}$ optical with three acoustics) are provided in supplementary document, Fig. S10 to Fig. S13. As it is depicted in Fig. 8 (a), directional behavior can be compared for each structure in separate band. For the acoustic modes, by approaching the Γ-point, the wavelength in the phase space becomes larger until reaching to infinity at Γ-point. Zero group velocity near the Γ-point reveals the 2D characteristic of diamane structures for ZA mode. Among the diamane structures, $C_2H$ shows the most similar behavior to graphene for ZA mode. For ZA mode, $C_4HCl$ shows a higher scattering rate, lower group velocity, and greater γ value, which can explain its lower contribution to the thermal conductivity. In contrast, $C_2H$ exhibit less scattering channels and a higher group velocity, which explain larger ZA contribution and thermal conductivity. Interestingly, the optical modes show a higher scattering rate (Fig. S11) and smaller weighted phase space (Fig. S13), which is the outcome of crossing of optical bands and narrow width for phonon dispersion.

For the group velocity (Fig. 8 (c)), $C_2H$ shows the highest phonon velocity among the diamane structures. $C_2Cl$, $C_4FCl$ and $C_2F$ show the lowest group velocity. However, the group velocity is the eminent parameter to determine the thermal conductivity of acoustic phonons contributions. For the scattering rate (Fig. 8 (d)), $C_2H$ shows the smallest amount of scattering among the diamane structures. However, for LOM in $C_2F$ and for only first optical mode in $C_2Cl$ structure, the high contribution can be explained well by its small amount of scattering rate (Fig. S11). In $C_2F$, however, $|\gamma|$ is smaller for ZA than for TA or LA; thus its contribution to thermal conductivity is the lowest in acoustic bands. For the Grüneisen parameter shown in Fig. S12, it is distinguishable which phonons in the q-space and from which bands show the strong correlation with lattice (strong anharmonicity). For Janus and non-Janus structures, respectively, $C_4HCl$ and $C_2Cl$ exhibit stronger anharmonicity (Fig. S12). Larger weighted phase spaces for graphene and diamane near the Γ-point represent the quadratic principle for 2D structures (ZA mode in Fig. 8 (f)), which causes more scattering channels to emit to other points in an absorption process. Similar to Grüneisen, for Janus and non-Janus structures, respectively, $C_4HCl$ and $C_2Cl$ have the highest amount of weighted phase space (channel to scatter) (Fig. S13).

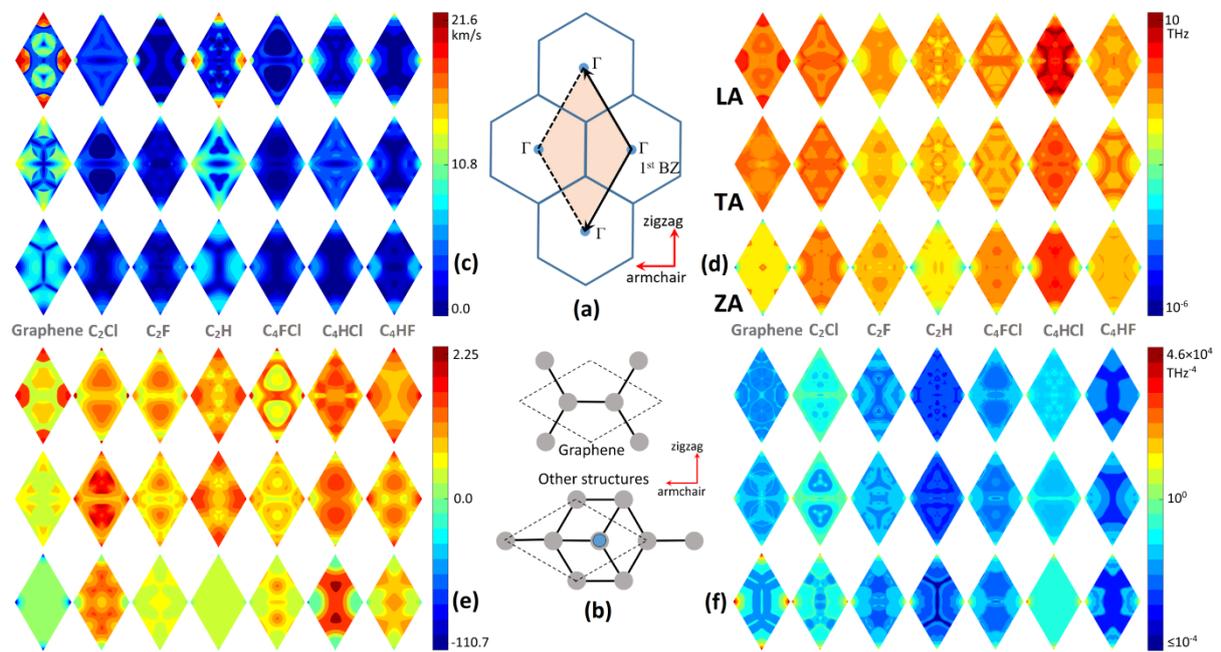

Fig. 8, Reciprocal lattice for the first Brillouin zone with illustration of zigzag and armchair directions (a), graphene and diamane structures (b), disparate acoustical contours for group velocity (c), scattering rate (d), Grüneisen parameter (e), weighted phase space (f). From left to right: graphene, $C_2Cl$, $C_2F$, $C_2H$, $C_4FCl$, $C_4HCl$, and $C_4HF$. From bottom to top: ZA, TA, and LA. Group velocity and Grüneisen parameter are of linear variation. Scattering rate and weighted phase space are of logarithmic variations.

Worthy to note that the monotonic increasing of the scattering rate (Fig. S7) and lower number for density of phonons with higher frequencies (Fig. S1), reveal that high-energy phonons show less population, shorter lifetime and lower contribution to heat transfer. In contrast, for low-energy phonons, the scattering rate is low and the density of phonons is high, which increases their heat transport in crystal. Fig. S14 demonstrates emission and adsorption processes over the q-space. For simplicity, this figure only shows which scattering process is dominated locally in the q-space. As it is expected, the low-energy phonons act as absorber, whereas high-energy phonons are emitter. Interestingly, graphene shows a different scattering behavior than those of diamane structures for TA and LA. Near the Γ-point, diamane structures are more phonon absorber than emitter, while for graphene, phonons are more emitter. Probably, this difference results from a large number of scattering channels for diamane structures and few bands in graphene. In addition to the mentioned reasons, emitter phonons (near the Γ-point for TA and LA modes) increases the phonon population of ZA (the most possible, and stable, band to be created) in graphene. On the contrary, the absorber phonons in the diamane structures keep LA, and TA, contribution in peer order of ZA, and thus increase the population of high-energy phonons resulting in more scattering and lower thermal conductivity. As follows from Fig. 5 for $C_2H$, acoustic modes and first three optical phonons (LOM) convincingly define the total thermal conductivity in this system. This can be explained by synergy of three factors: first, for these low frequency bands the absorption is the dominating process near the Γ-point (Fig. S14). Second, these aforementioned bands are dispersed far from each other and from higher optical modes, with less crossing bands, which result in fewer scattering channels (Fig. 3). Third, showing lower scattering rates in low-energy phonons with high number density of phonons (Fig. S7). For $C_4ClH$, highest optical band has both locally dominating absorption and emission (Fig. S14). This difference among the diamane structures is due to the mound-like dispersion of highest band in figure, which is depicted in Fig. S15 (a). Near the Γ-point of highest optical mode, the energy is locally low, which include this area of q-space into the absorption process with only low acoustic branches to obtain the higher energy level. All diamane structures, specifically $C_4HCl$, have higher scattering rate in their high-energy phonons, which means that these phonons are active in the lattice.

## 4. Summary

In this work, we explored electronic properties and thermal conductivity of non-Janus $C_2H$, $C_2F$ and $C_2Cl$ and Janus $C_4HF$, $C_4ClH$ and $C_4ClF$ dimanes, a novel class of carbon-based 2D materials. These novel 2D systems show very diverse electronic properties, with electronic band gaps ranging from 0.86 to 5.68 eV. On the basis of HSE06 method, the band gap of $C_2H$, $C_2F$, $C_2Cl$, $C_4HF$, $C_4HCl$ and $C_4FCl$ diamanes are predicted to be 3.86, 5.68, 2.42, 4.17, 0.86, and 2.05 eV, respectively. We employed a computationally robust method to estimate the thermal conductivity of diamane nanosheets. In this approach, thermal properties were acquired by the full iterative solutions of the Boltzmann transport equation, with the accelerated calculation of the anharmonic force constants using machine-learning interatomic potentials. The room temperature lattice thermal conductivity of graphene and $C_2H$, $C_2F$, $C_2Cl$, $C_4HF$, $C_4HCl$ and $C_4FCl$ diamane monolayers are estimated to be 3636, 1145, 377, 146, 454, 244 and 196 W/mK, respectively. Our extensive analysis highlights the substantial role of functional groups on the electronic and thermal conduction responses of diamane nanosheets, which may serve as valuable guide for future studies.

Acknowledgment

B. M. and X. Z. appreciate the funding by the Deutsche Forschungsgemeinschaft (DFG, German Research Foundation) under Germany's Excellence Strategy within the Cluster of Excellence PhoenixD (EXC 2122, Project ID 390833453). E.V.P. and A.V.S. were supported by the Russian Science Foundation (Grant No 18-13-00479).

# Supporting Information

Atomic lattices in VASP POSCAR format:
C2H
```
  1.00000000000000
    2.5177450641725931    0.0000000000000000    0.0000000000000000
    1.2588725320862970    2.1804311856098328    0.0000000000000000
    0.0000000000000000    0.0000000000000000   20.0000000000000000
   C   H
    4   2
Direct
 0.3333333690000018  0.3333333579999973  0.5115092844652835
 0.3333333690000018  0.3333333579999973  0.5891598134335823
 0.6666667380000035  0.6666667160000017  0.6133715076031194
 0.0000000000000000  0.0000000000000000  0.4872951761247748
 0.6666667380000035  0.6666667160000017  0.6690956150771967
 0.0000000000000000  0.0000000000000000  0.4315686012960427
```

C2F
```
1.00000000000000
    2.5463561367547811    0.0000000000000000    0.0000000000000000
    1.2731780683773919    2.2052091008911048    0.0000000000000000
    0.0000000000000000    0.0000000000000000   20.0000000000000000
   C   F
    4   2
Direct
 0.3333333690000018  0.3333333579999973  0.5116838973901991
 0.3333333690000018  0.3333333579999973  0.5889852005086667
 0.6666667380000035  0.6666667160000017  0.6141003339282705
 0.0000000000000000  0.0000000000000000  0.4865663497996238
 0.6666667380000035  0.6666667160000017  0.6826042271496320
 0.0000000000000000  0.0000000000000000  0.4180599892236074
```

C2Cl
```
1.00000000000000
    2.7340933967969931    0.0000000000000000    0.0000000000000000
    1.3670466983984979    2.3677943384027929    0.0000000000000000
    0.0000000000000000    0.0000000000000000   20.0000000000000000
   C   Cl
    4   2
Direct
 0.3333333690000018  0.3333333579999973  0.5120422083239191
 0.3333333690000018  0.3333333579999973  0.5886268895749467
 0.6666667380000035  0.6666667160000017  0.6138103948152747
 0.0000000000000000  0.0000000000000000  0.4868562889126196
 0.6666667380000035  0.6666667160000017  0.7007907232683053
 0.0000000000000000  0.0000000000000000  0.3998734931049270
```

```
C4HF
   1.00000000000000
     2.5320000000000000    0.0000000000000000    0.0000000000000000
     1.2660000000000000    2.1927763219999998    0.0000000000000000
     0.0000000000000000    0.0000000000000000   20.0000000000000000
   C  H  F
   4  1  1
Direct
  0.3333333690000018  0.3333333579999973  0.5117201190616868
  0.3333333690000018  0.3333333579999973  0.5892935866974980
  0.6666667380000035  0.6666667160000017  0.6135290347418978
  0.0000000000000000 -0.0000000000000000  0.4867172839872369
  0.6666667380000035  0.6666667160000017  0.6691435145861054
  0.0000000000000000 -0.0000000000000000  0.4180372814405819

C4HCl
   1.00000000000000
     2.6402000000000001    0.0000000000000000    0.0000000000000000
     1.3201000000000001    2.2864802709999998    0.0000000000000000
     0.0000000000000000    0.0000000000000000   20.0000000000000000
   C  Cl  H
   4  1  1
Direct
  0.3333333690000018  0.3333333579999973  0.5061190328188515
  0.3333333690000018  0.3333333579999973  0.5833561947399322
  0.6666667380000035  0.6666667160000017  0.6082963127508044
 -0.0000000000000000 -0.0000000000000000  0.4822349375344532
  0.6666667380000035  0.6666667160000017  0.6953319242749868
 -0.0000000000000000 -0.0000000000000000  0.4266615958809500

C4FCl
   1.00000000000000
     2.6507000000000001    0.0000000000000000    0.0000000000000000
     1.3253500000000000    2.2955735380000002    0.0000000000000000
     0.0000000000000000    0.0000000000000000   20.0000000000000000
   C  Cl  F
   4  1  1
Direct
  0.3333333690000018  0.3333333579999973  0.5086599621585393
  0.3333333690000018  0.3333333579999973  0.5857377739625989
  0.6666667380000035  0.6666667160000017  0.6108559294092473
  0.0000000000000000  0.0000000000000000  0.4837816977342655
  0.6666667380000035  0.6666667160000017  0.6976863590691877
  0.0000000000000000  0.0000000000000000  0.4152782756661466
```

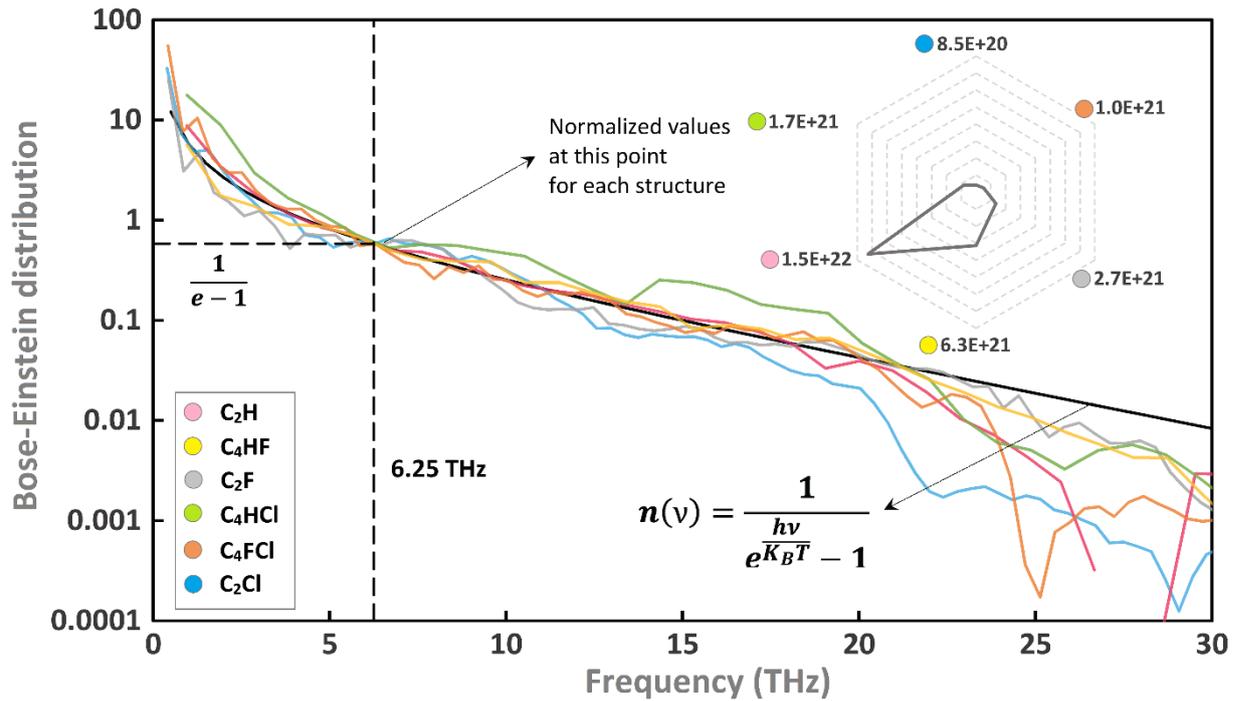

**Fig. S1**, Bose-Einstein distribution of diamane structures for room temperature are normalized to the value of inset hexagonal chart, which attributed to specified diamane structure.

As discussed in the main paper, the related frequency of room temperature is 6.25 THz. This figure reveals the relation of Bose-Einstein distribution and individual thermal conductivities as a function of frequency, which is normalized by their value when frequency selected as 6.25 THz. The higher thermal conductivity happened if two reasons guaranteed, first, the number of phonons passed through the unit area in specific time were high, or, second, the energy of passed phonons were high. For the fixed room temperature comparison, the first reason is the answer and these normalized value (inset hexagonal chart) associated to the number density of phonons transportation, in which $C_2H$ structure shows the highest value.

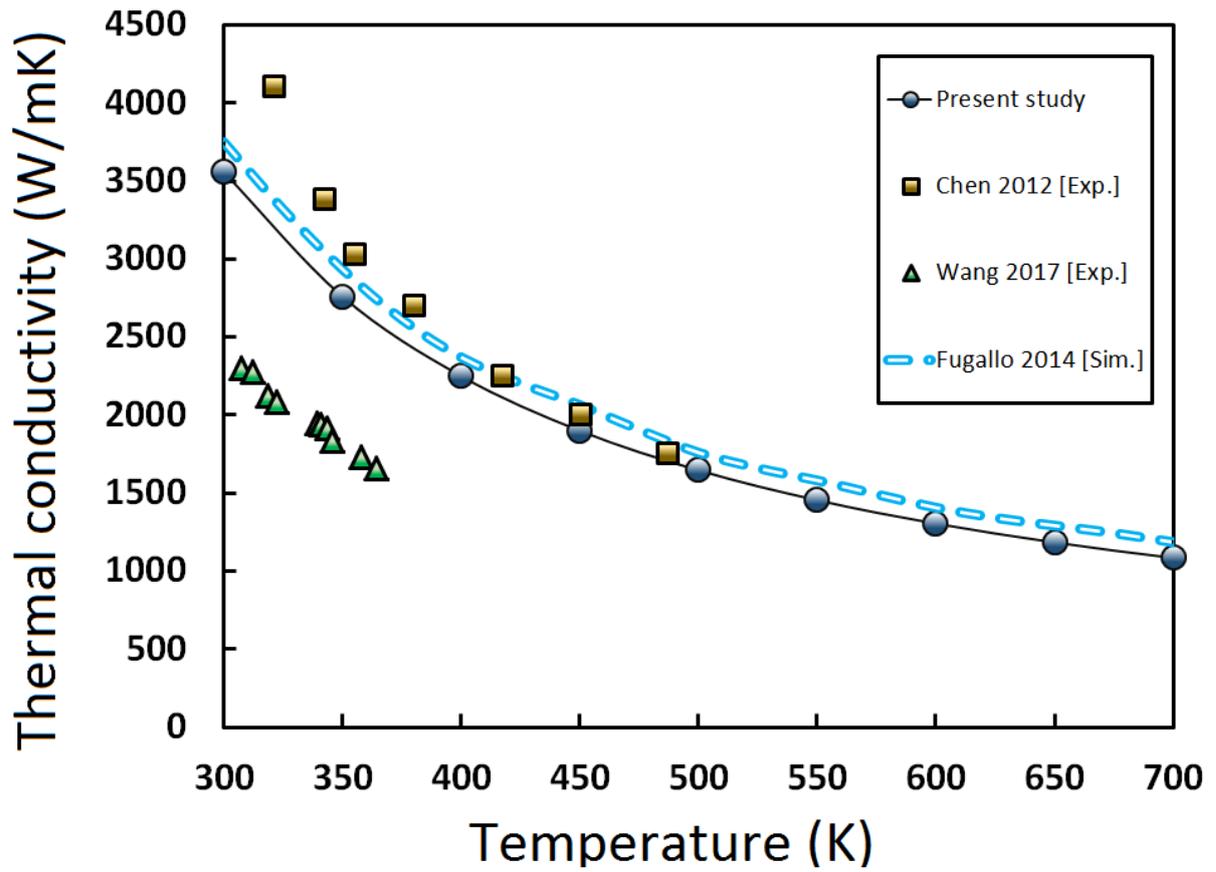

Fig. S2, Thermal conductivity of graphene as a function of by the present study on the basis of MTP results and previous works [1–5].

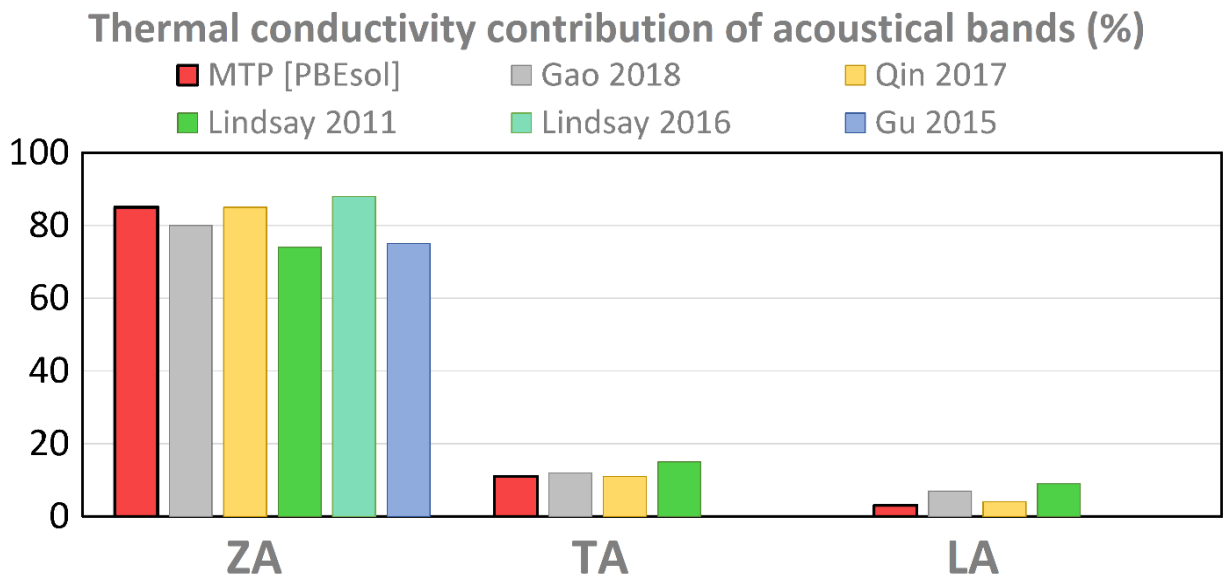

Fig. S3, Contribution of acoustic modes on the thermal conductivity of graphene by the present study on the basis of MTP results compared with previous works [6–10].

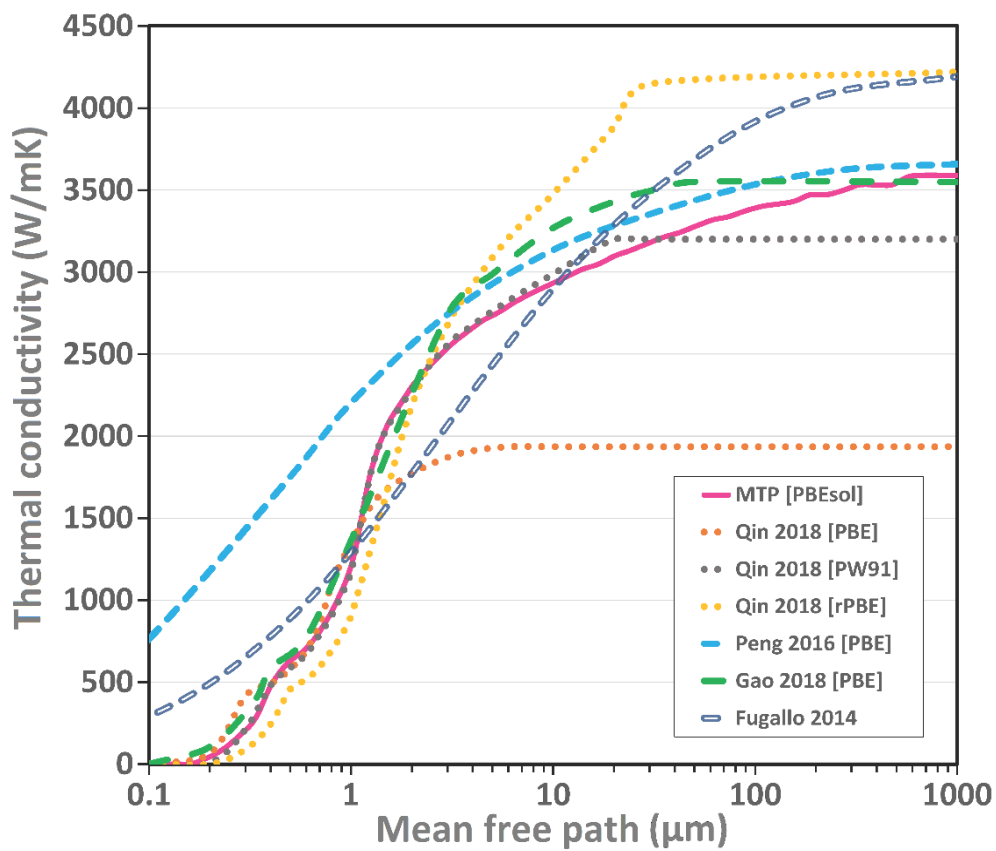

**Fig. S4**, Cumulative thermal conductivity graphene by MTP with previous full DFT works [1,6,11,12].

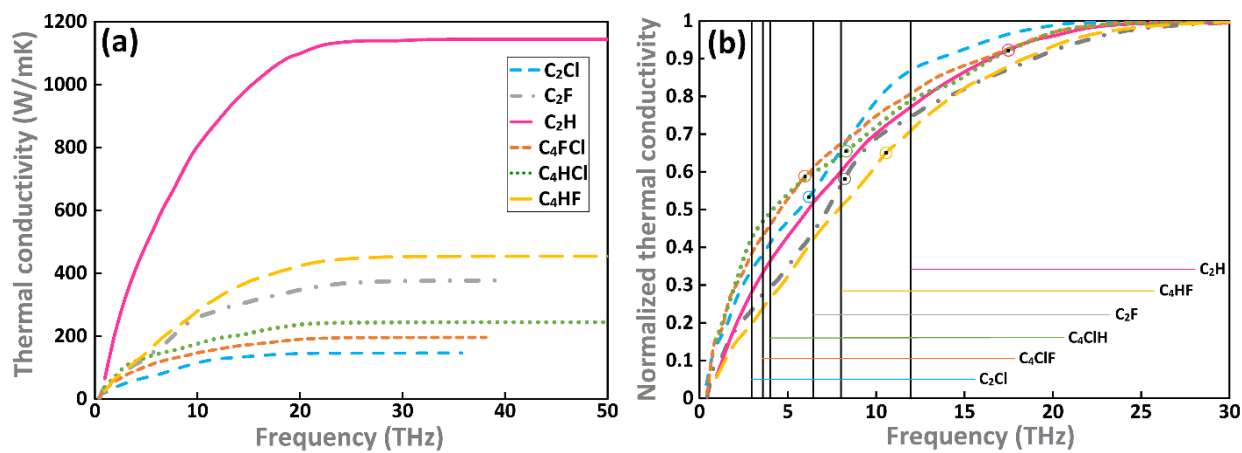

**Fig. S5**, cumulative thermal conductivity as a function of frequency (a) and its normalized (b) for diamane structure at room temperature. For right graph (b), vertical lines pinpoint the start point of optical modes for each structure, and, dotted circles are contribution of acoustical bands (ZA, TA, and LA) to the thermal conductivity.

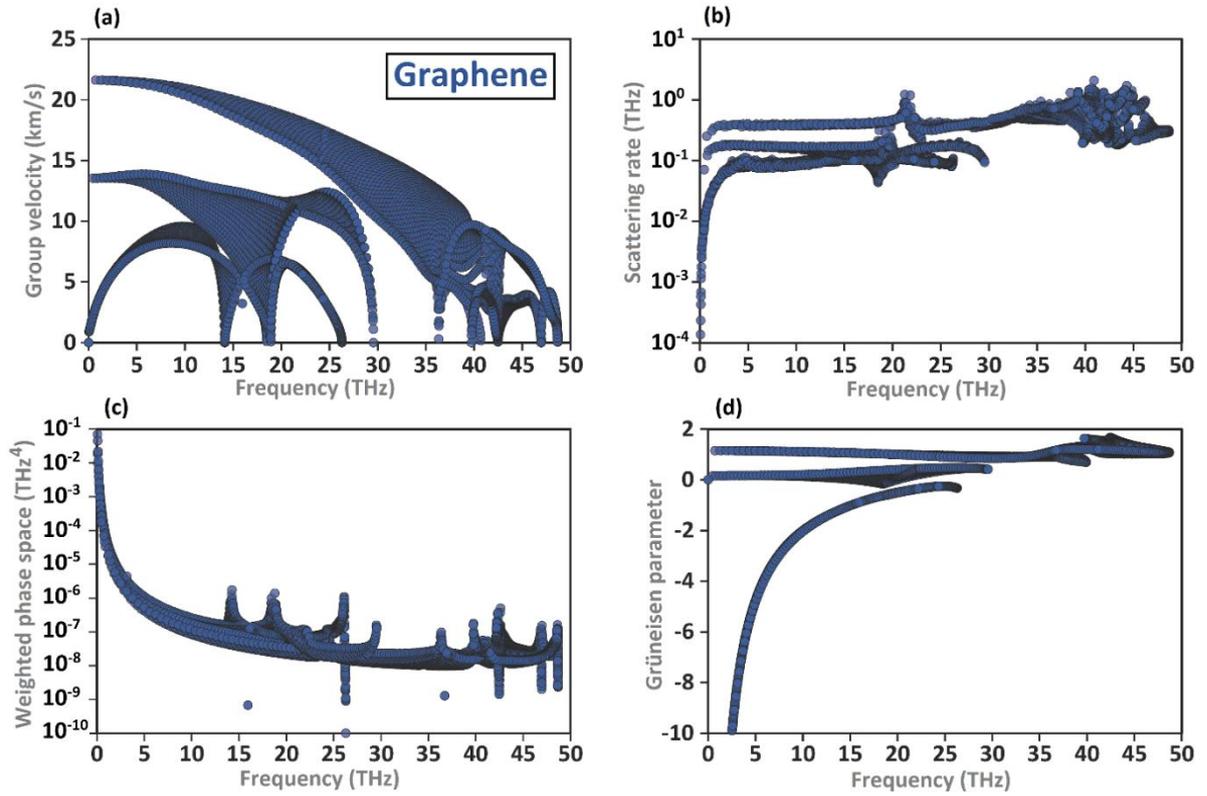

**Fig. S6**, (a) group velocity, (b) scattering rate, (c) weighted phase space and (d), Grüneisen parameter of graphene as a function of frequency.

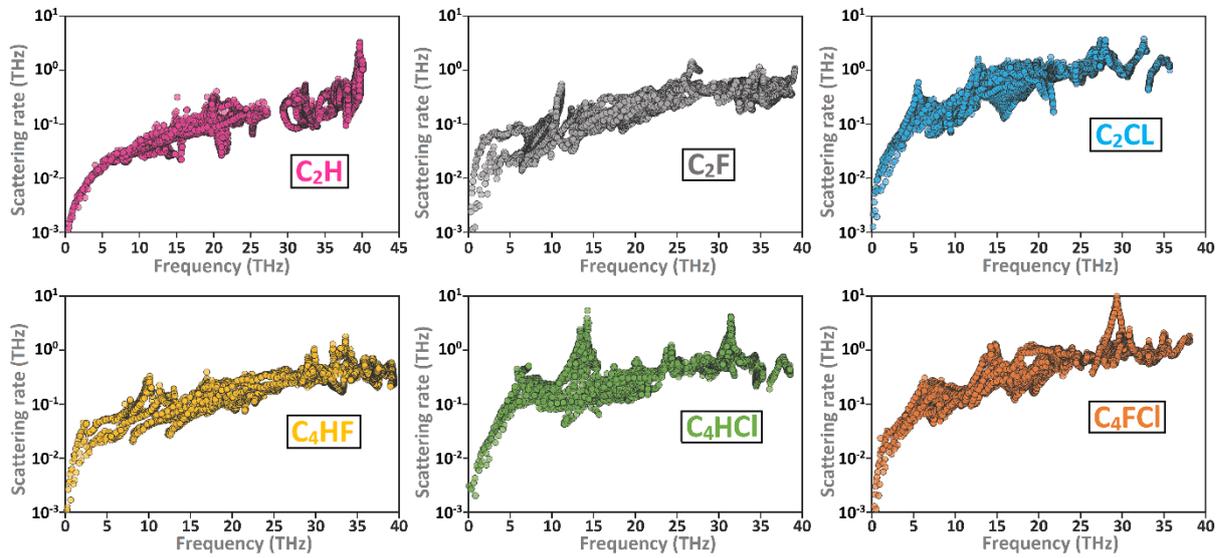

**Fig. S7**, phonon scattering rate as a function of frequency for diamane structures

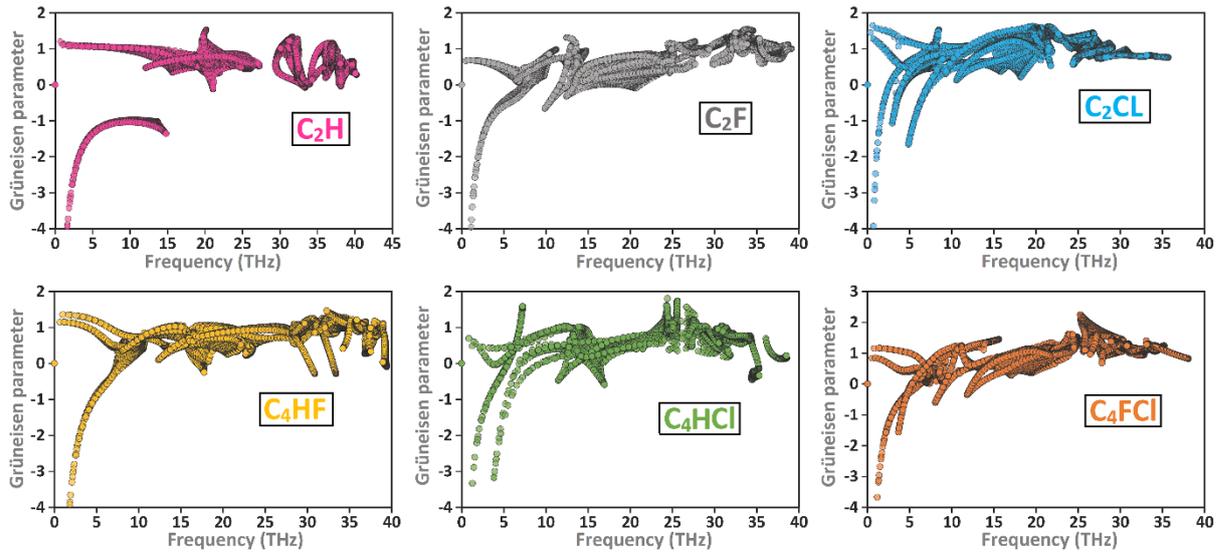

**Fig. S8**, Grüneisen parameter as a function of frequency for diamane structures

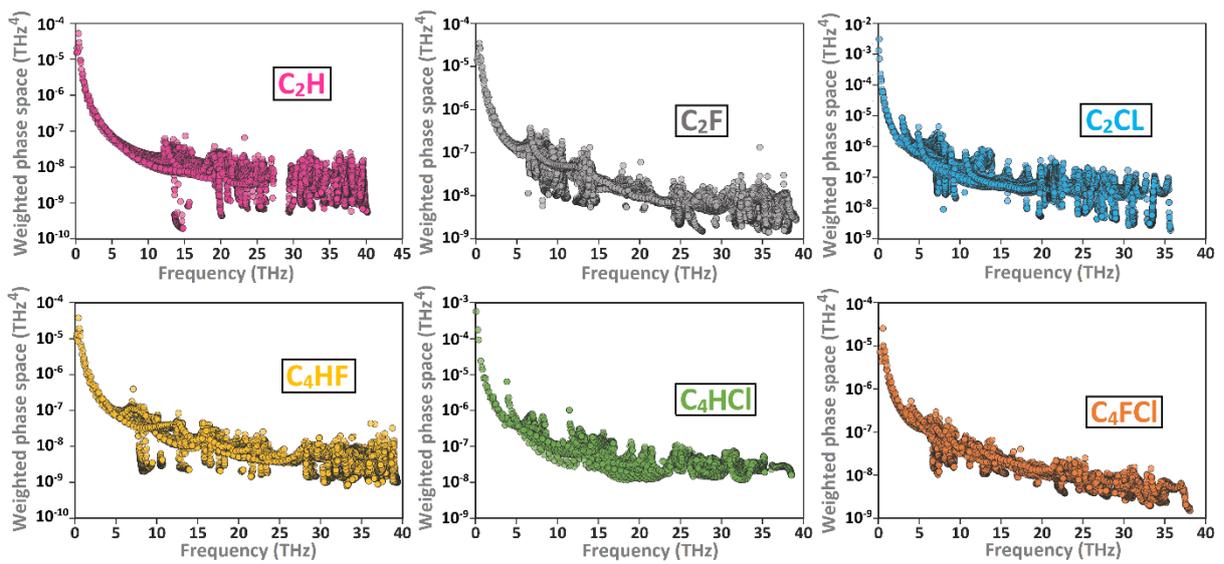

**Fig. S9**, weighted phase space as a function of frequency for diamane structures

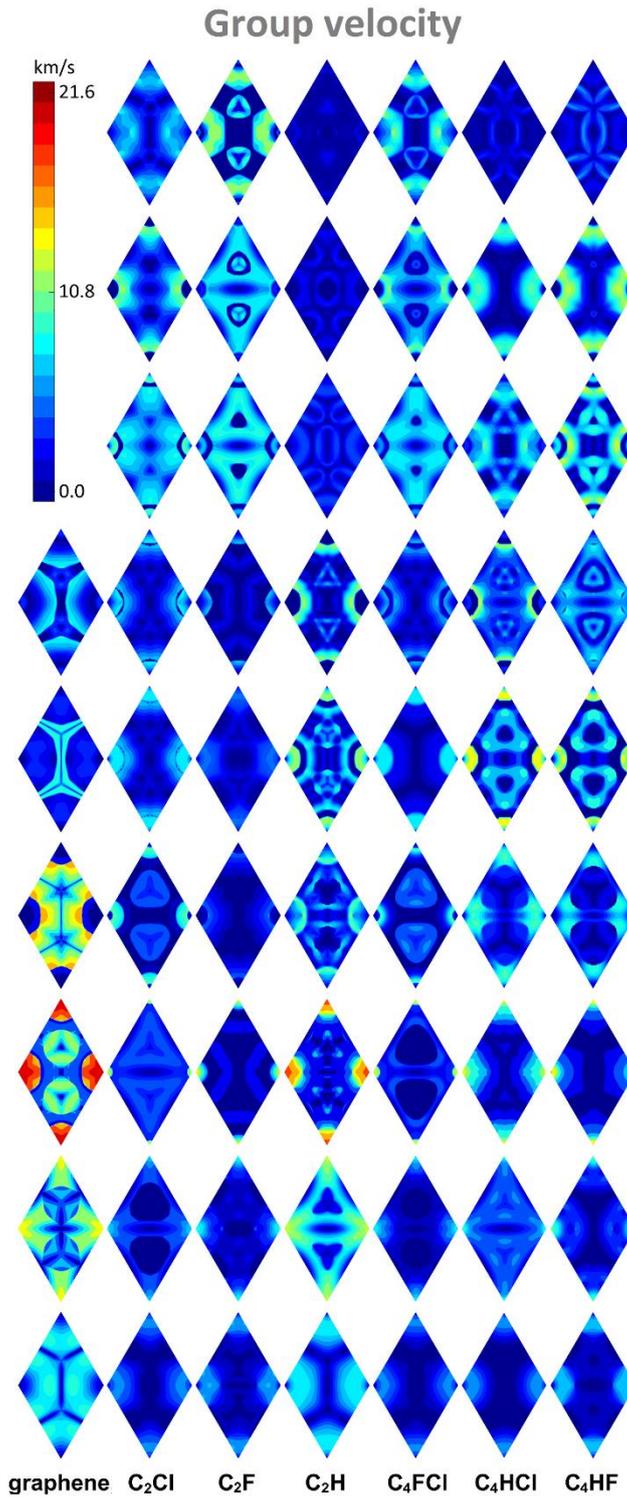

**Fig. S10**, Mapping of the group velocity over the reciprocal lattice for acoustical and optical (up to 6[th]) modes. From bottom to top results belong to ZA, TA, LA and six first optical modes, respectively.

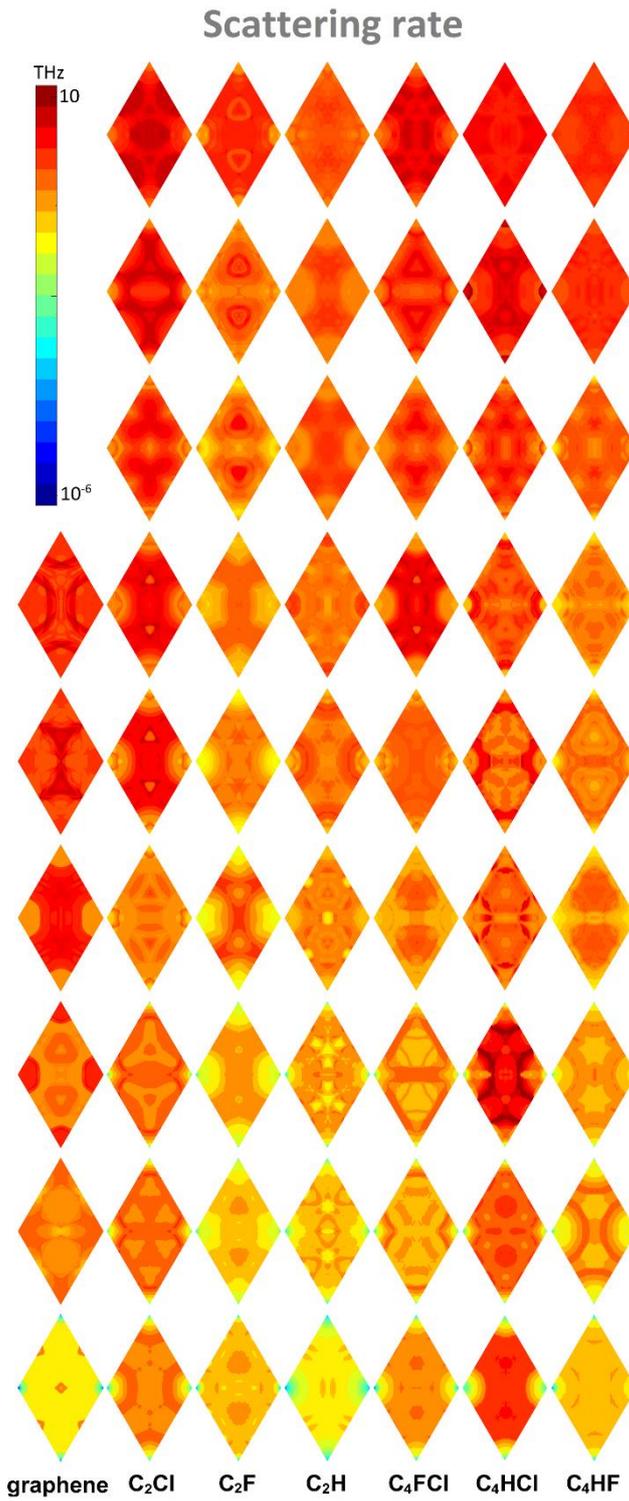

**Fig. S11**, Mapping of the scattering rate over the reciprocal lattice for acoustical and optical (up to 6[th]) modes. From bottom to top results belong to ZA, TA, LA and six first optical modes, respectively.

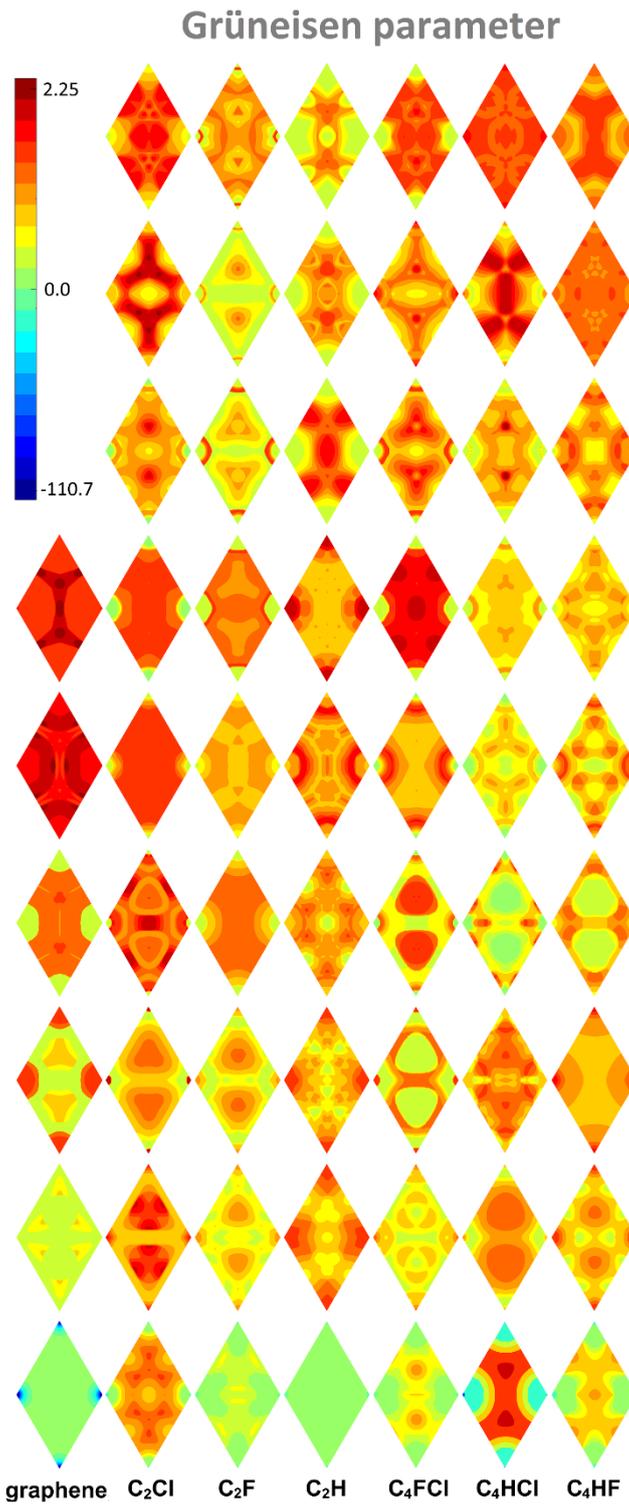

**Fig. S12**, Mapping of the Grüneisen parameter over the reciprocal lattice for acoustical and optical (up to 6[th]) modes. From bottom to top results belong to ZA, TA, LA and six first optical modes, respectively.

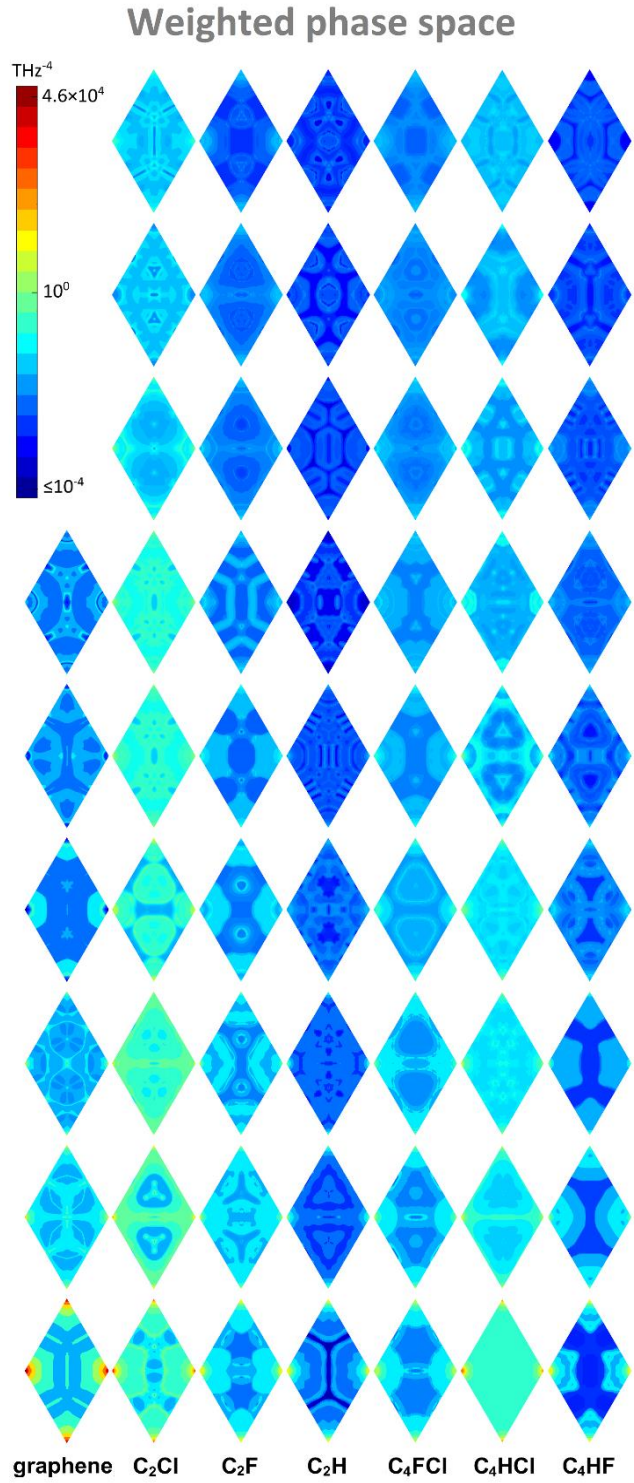

Fig. S13, Mapping of the weighted phase space over the reciprocal lattice for acoustical and optical (up to 6[th]) modes. From bottom to top results belong to ZA, TA, LA and six first optical modes, respectively.

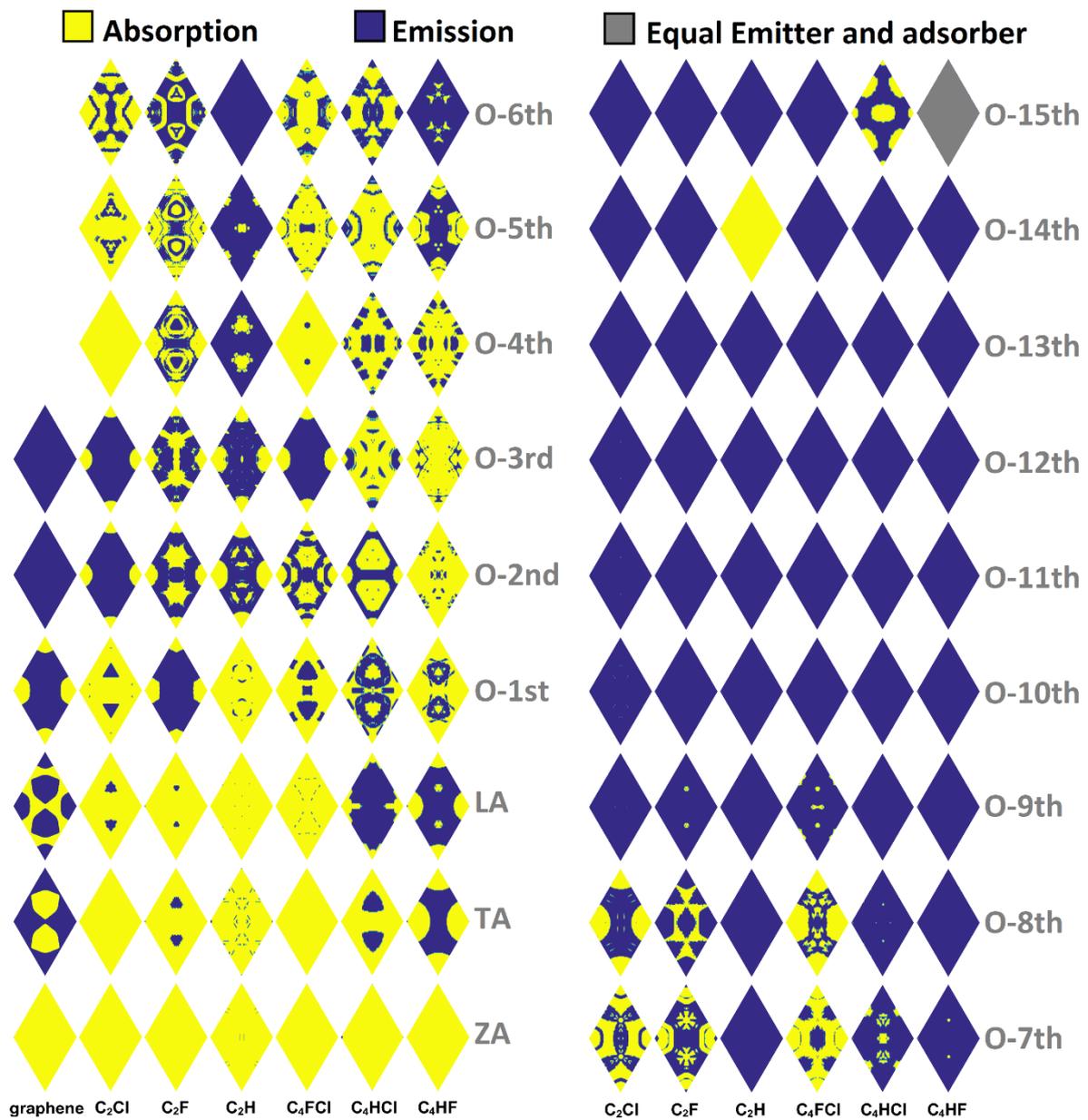

**Fig. S14**, Contour of governed process over q-space. Blue area refers that emission process is higher than absorption, and reversely, yellow sections refer to governing of absorption process. Grey color represents evenly emission and adsorption process.

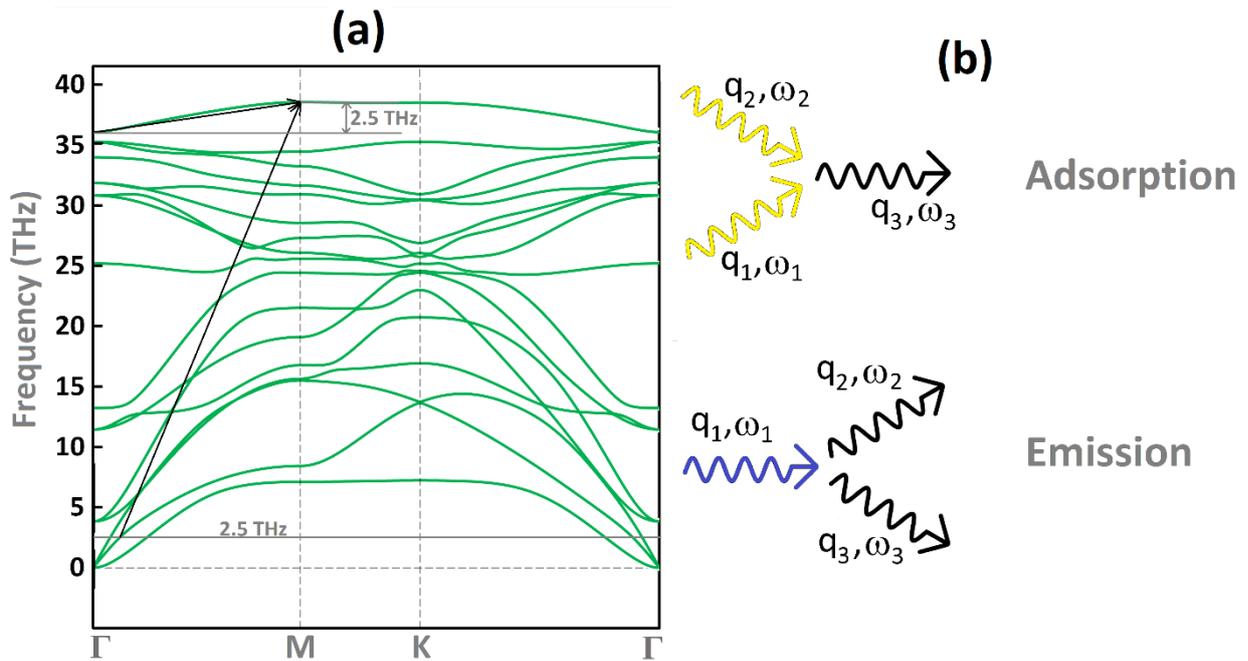

Fig. S15, (a) Graphical explanation for highest-energy phonon adsorption process in Fig. S14 (O-15[th]). (b) Graphical illustration of adsorption and emission mechanism in three-phonon scattering.